\documentstyle[times,epsfig]{jaa}
\begin{document}
\title[]{Generalized Langevin equation description of the stochastic oscillations of general relativistic disks}
\author[]%
       {Chun Sing Leung$^1$\thanks{E-mail:csinleung@polyu.edu.hk}, Gabriela Mocanu $^2,^3$\thanks{Email:gabriela.mocanu@ubbcluj.ro}, Tiberiu Harko$^4$\thanks{E-mail:t.harko@ucl.ac.uk}\\
        $^1$Department of Applied Mathematics, Polytechnic University, Hong Kong\\
         $^2$Faculty of Physics,
          Babes-Bolyai University, Cluj-Napoca, Romania\\ $^3$ Faculty of Automation and Computer Science, Technical University, Cluj Napoca, Romania \\
         $^4$ Department of Mathematics, University College London,  United Kingdom }
\maketitle
\label{firstpage}
\begin{abstract}
We consider a description of the stochastic oscillations of the general relativistic accretion disks around compact astrophysical objects based on the generalized Langevin equation, which accounts for the general retarded effects of the frictional force, and on the  fluctuation-dissipation
theorems. The vertical displacements, velocities and luminosities of the stochastically perturbed disks are explicitly
obtained for both the Schwarzschild and the Kerr cases. The Power Spectral Distribution of the luminosity it is also obtained, and it is shown that it has non-standard values. The theoretical predictions of the model are compared with the observational data for the luminosity time variation of the BL Lac S5 0716+714 object.
\end{abstract}

\section{Introduction}
\label{sec:intro}

In an astrophysical environment, compact general relativistic objects like black holes or neutron stars are
often surrounded by an accretion disk. Accretion disks around compact objects have become standard models for a number of astrophysical phenomena like X-ray binaries or active galactic nuclei. The disks are usually considered as being composed of massive test particles that move in the gravitational ﬁeld of the central compact object.  It was also suggested that the Intra Day Variability of AGNs can also be related to the vertical oscillations of the accretion disks around black holes (Mineshige et al. 1994;  Leung et al. 2011).

The study of the general relativistic oscillations of thin accretion disks around compact astrophysical objects interacting with the surrounding medium through non-gravitational forces was initiated in Leung et al. (2011) and Harko \& Mocanu (2012). The interaction of the accretion disk with the external medium (a thermal bath) was
modeled via a friction force, and a random force, respectively. By taking into account the presence of a viscous dissipation and of a stochastic force the dynamics of the stochastically perturbed disks can be formulated in terms of a general relativistic Langevin equation.

It is the purpose of the present paper to extend the analysis initiated in Leung et al. (2011) and Harko \& Mocanu (2012)  for the description of the oscillations of the particles in the accretion disks around compact objects in contact with an external heat bath. We assume that the oscillations of the disk, assumed to interact with an external medium (the heat bath), are described by the generalized Langevin equation, introduced by Kubo (1964).  As an application of our model we consider the possibility that stochastic disk oscillations may be the source of the Intra Day Variability observed in Active Galactic Nuclei type compact objects.

\section{The generalized Langevin equation for stochastic oscillations of accretion disks}

 We consider that the particles in the disk are in contact with
an isotropic and homogeneous external medium. The interaction of the particles
with the cosmic environment is described by a friction force, and a random force. The vertical oscillations of the disk under the influence of a stochastic force $\xi ^z $ can be described by
 the generalized Langevin equation, introduced by Kubo (1964), given by
\begin{equation}\label{genLang}
\frac{d^2\delta z}{dt^2}+\int_0^t{\gamma (t-\tau) \frac{d\delta z (\tau )}{d\tau}}d\tau+ V'(\delta z)=c^2 \frac{\xi ^z(t)}{M_D},
\end{equation}
where
\begin{equation}
\gamma (\vert t-\tau \vert) = \frac{\alpha c}{\tau _d} \exp \left \{ -\frac{\vert t-\tau \vert}{\tau _d} \right \},
V (\delta z) = \frac{c^2 \omega _\perp ^2}{2} \left ( \delta z \right)^2,
\langle \xi ^z (t) \xi ^z (\tau) \rangle = \frac{1}{\beta} \gamma (t-\tau),
\end{equation}
where $\omega _{\perp}^2$ is the disk oscillation frequency, $M$ is the mass of the central object, $M_D$ is the mass of the disk, and $\alpha $, $\beta $ and $\tau _d$ are constants.
By introducing a set of dimensionless parameters $(\theta, \sigma, \delta Z)$ defined as
$t=\tau _d\theta$, $\tau =\tau _d \sigma $, $\rho =nGM/c^2$, $n\geq 6$, and $\delta z=\left(c^2\tau _d^2/M_D\right)  \overline{q}$,
the equation of motion of the stochastically oscillating disk can be written as
\begin{equation}
\frac{d^2\overline{q}}{d\theta ^2}+a\int_0^{\theta }{e ^{-\vert \theta-\sigma \vert} \frac{d\overline{q} (\sigma )}{d\sigma}d\sigma }+b \overline{\omega} _\perp ^2 \overline{q}=\overline{\xi} (\theta ),\label{eq:GLEnoDim}
\end{equation}
where we have denoted $a=\alpha c\tau _d$, $b=c^2\tau _d^2/M^2$, $\xi ^z = \left(1/M_D\right)\overline{\xi}(\theta)$, and
$\omega _\perp  M =\overline{\omega} _ \perp $, respectively.
The energy of the disk is the sum of kinetic plus potential energy$
E = (1/2) \left [ \left(d \delta z/dt \right )^2 + V(\delta z)\right]$,
and the output luminosity, given by the time variation of the energy,
$L = - dE/dt$ can be written in dimensionless form as
$L = \left(c^4 \tau _d/M_d^2\right) \overline{L}$, where
\begin{equation}
\overline{L} = - \frac{d\overline{q}}{d\theta} \left ( \frac{d^2 \overline{q}}{d\theta ^2} + b \overline{\omega}_\perp ^2 \overline{q} \right ).
\end{equation}
The dimensionless noise $\overline{\xi}$ has the properties
$\langle \phi (\theta) \phi (\sigma) \rangle = A \delta (\theta - \sigma)$, where $A$ is a constant. The generalized Langevin equation is numerically integrated by using the algorithm introduced in Hershkovitz (1998).
The behaviour of the dimensionless luminosity is represented, for the case of the Schwarzschild and Kerr black holes, respectively,  in Figs.~1.

\begin{figure}[!h]\label{Leung}
\includegraphics[height=1.4in, width = 2.2in]{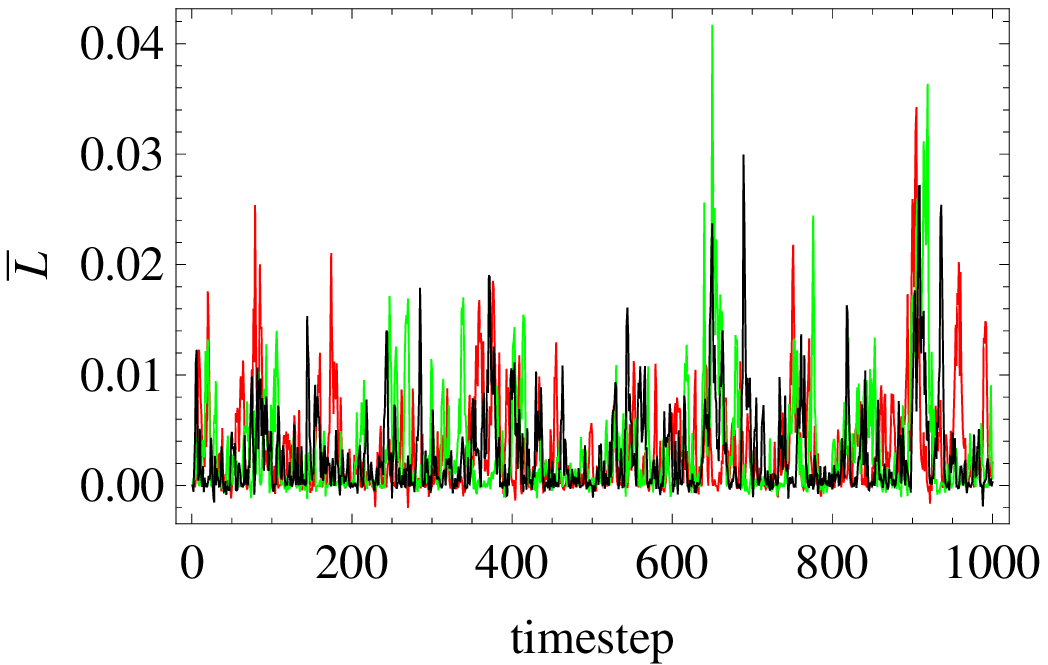}
\includegraphics[height=1.4in, width = 2.2in]{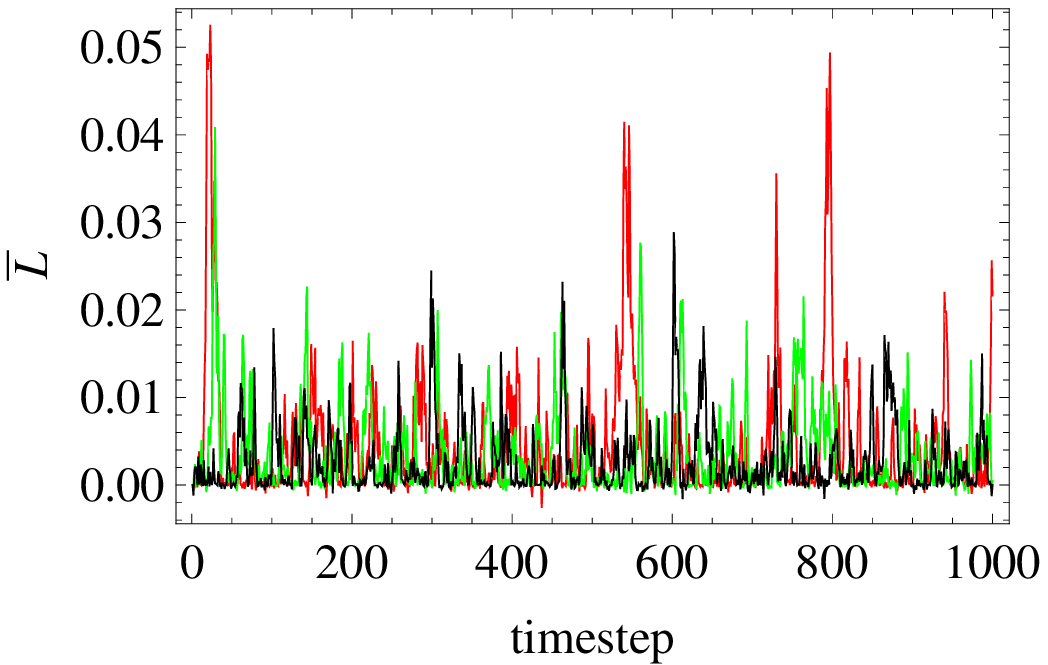}
\caption{Luminosity of a stochastically oscillating accretion disk  around a Schwarzschild black hole (left figure), for $A = 0.0015$ and $a=0.03$ (red curve), $a = 0.1$ (green curve), and $a=0.001$ (black curve), respectively, and for a Kerr black hole (right figure), for $A = 0.0015$ and $a=0.03$ (red curve), $a = 0.1$ (green curve), and $a=0.001$ (black curve), respectively.}
\end{figure}

 The PSDs (Power Spectral Distributions)  of the luminosities  are shown in Figs.~2. The PSD curves were obtained by using the .R software (Vaughan 2010). The black curve represents the points produced by taking the PSD of the simulated light curve, the blue curve represents the best fit, and the red curve represents the fit according to a power-law of the PSD, $P\sim f^{-\alpha}$, respectively.

\begin{figure}[!h]\label{Leung2}
\includegraphics[height=1.4in, width = 2.2in]{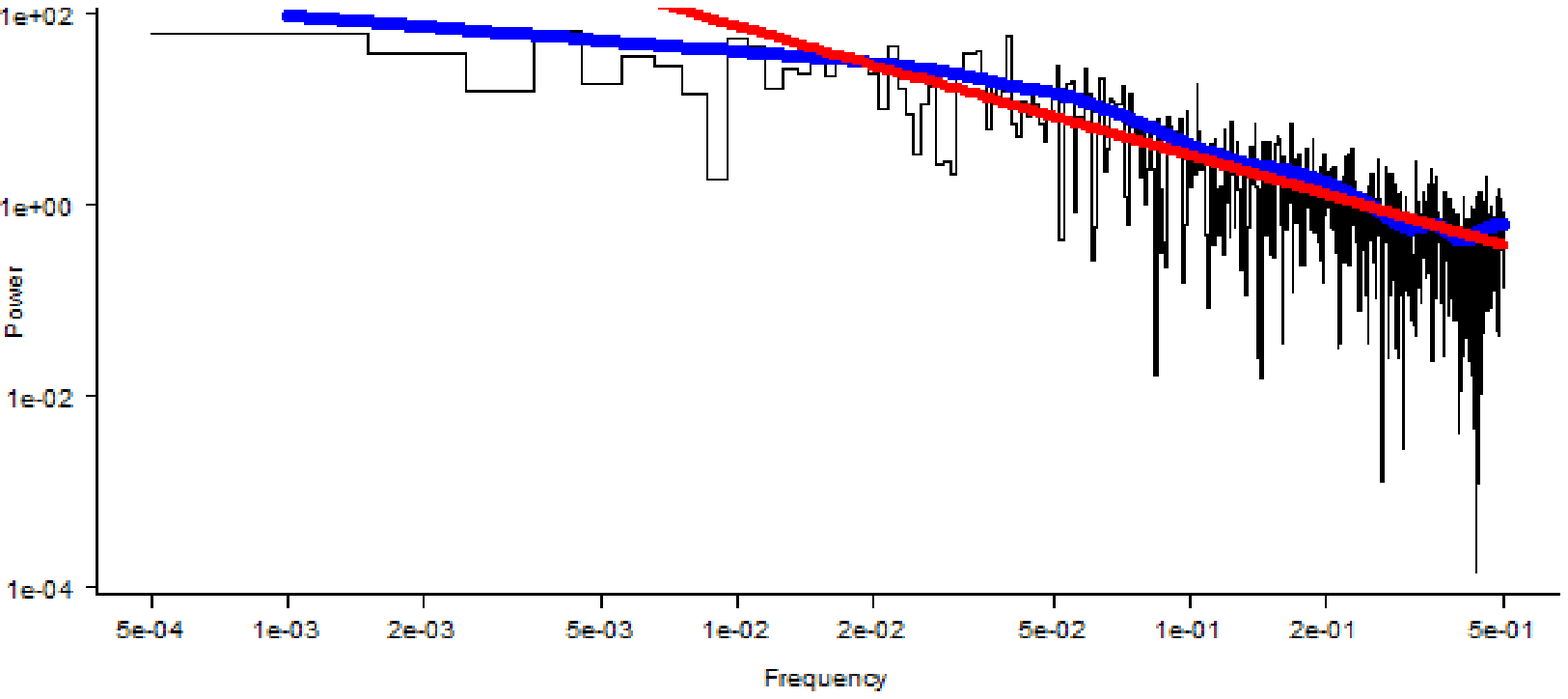}
\includegraphics[height=1.4in, width = 2.2in]{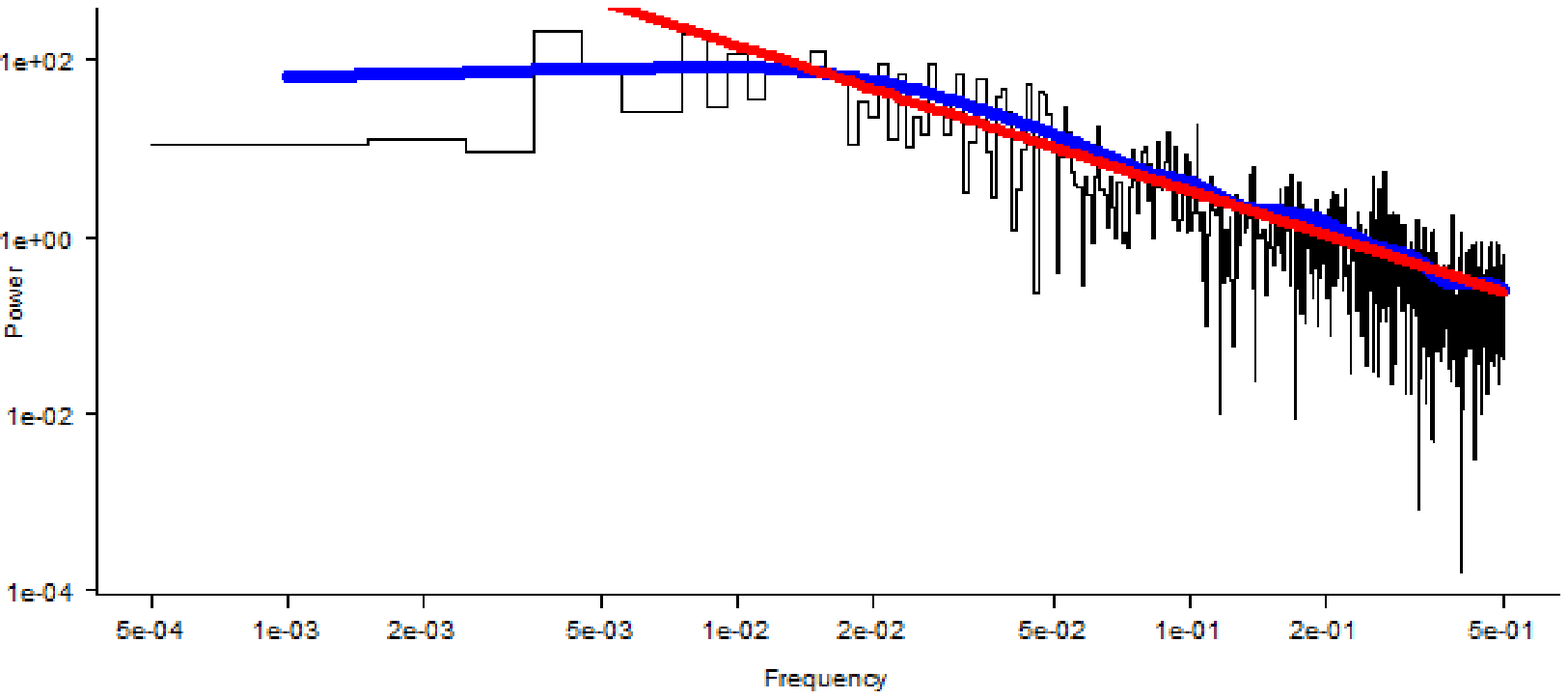}
\caption{PSDs of the luminosities for the Schwarzschild case (left figure), and for the Kerr case (right figure), respectively. The blue curve represents the best fit of the data points, while  the red curve represents the fit according to a power-law, $P\sim f^{-\alpha}$.}
\end{figure}

 \section{Comparison with AGN IDV data}

For BL Lac S5 0716+714, an AGN with  $M \in [10^{7.68}M_\odot,$ $ 10^{8.38}M_\odot]$ (Fan et al. 2011), a PSD analysis performed on IDV data in the BVRI bands has shown that $\alpha$ varied between approximately $-1.5$ and $-2.7$ for that set of data (Mocanu \& Marcu 2012). Most of the light curves from that observational set have shown a spectral slope in the interval $1.5$ to $2$. The model presented in this work produces light curves which show spectral slopes within the second most populated interval, the $1.0$ to$1.5$ for modest rotation parameters. If the rotation parameter is increased, there is a specific radial distance that within this framework can produce slopes in the first interval. This radius should be placed between approximately $4R_S$ and $6R_S$, where $R_S=2GM/c^2$ is the Schwarzschild radius,  which is in remarkable agreement with theory and observations regarding the optical spectrum.
The model presented in this work is thus consistent with observational data, from the point of view of the imprint of the underlying light generating process on the PSD signature. This supports the idea that perturbations due to nontrivially correlated noise can consistently explain the statistical properties of the observed IDV light curves.

\section{Discussions and final remarks}

In the present paper we have investigated the stochastic fluctuations of the accretion disks around compact general relativistic objects due to the presence of a colored background noise. In this case the vertical oscillations of the disk are described by the generalized Langevin equation (Kubo 1964) in the presence of the general relativistic gravitational potential. In our approach we have assumed that the motion of the disk is non-relativistic in the sense that the velocity $\delta z/dt$ of the particles is much smaller than the speed of light, $\delta z/dt<<c$. We have investigated the behavior of oscillations in both the Schwarzschild and Kerr black hole cases. As an astrophysical application of our model we have considered the possibility of the description of the IDV phenomenon in the framework of the present model.

{\bf Acknowledgments.}  G.M is supported by POSDRU/107/1.5/S/76841.

\end{document}